\documentstyle[prl,aps,twocolumn,psfig,floats]{revtex}

\begin{document}
\title{Critical behavior of the $S=3/2$ antiferromagnetic Heisenberg chain}

\author{K. Hallberg$^{a}$, X.~Q.~G. Wang$^{a}$, P. Horsch$^{b}$ and 
A. Moreo$^{c}$}

\address{
$(a)$ Max-Planck-Institut f\"{u}r Physik komplexer Systeme, Bayreuther Str.~40,
D-01187 Dresden (Germany) \\
$(b)$ Max-Planck-Institut f\"{u}r Festk\"{o}rperforschung,
Heisenbergstr.~1, D-70569 Stuttgart (Germany) \\
$(c)$  Department of Physics. National High Magnetic Field Laboratory and
MARTECH. Florida State University, Tallahassee, Florida 32306, (U.S.A.)
\centerline{\rm (February 29th. 1996)}
\parbox[t]{14truecm}{\small
~\\
       Using the density-matrix renormalization-group technique we
study the long-wavelength properties of the spin $S=3/2$ 
nearest-neighbor Heisenberg chain. 
We obtain an accurate value for the spin velocity $v=3.87\pm 0.02$,
in agreement with experiment.
Our results show conclusively that the model
belongs to the same universality class as the $S=1/2$ Heisenberg chain,
with a conformal central charge $c=1$ and critical exponent
$\eta=1$. 
}}

\maketitle



The study of one-dimensional spin-S Heisenberg Hamiltonians has received
considerable attention in recent years. Their properties are important to
experimentalists because they describe a number of materials with
magnetic ions arranged in chains\cite{exp1,exp2}. The first quasi-1D 
antiferromagnets with spin 3/2 which found
experimental interest are $CsVCl_3$\cite{itoh95} and 
$AgCrP_2S_6$\cite{mutka95}.
For theorists the spin
models are among the simplest Hamiltonians where quantum fluctuations are
crucial. However, only the special case S=1/2 can be solved exactly\cite{bethe}. 
In this case Luther and Peschel\cite{luther} found that the spectrum is
gapless, there is no magnetic ordering in the ground state, and the spin
correlation function $\omega(l)$ 
decays according to a power law $|\omega(l)| \sim 1/l^{\eta}$ with
$\eta=1$. Later Haldane\cite{haldane} proposed that
for half-odd-integer (h.o.i.) spin the Heisenberg Hamiltonian 
would behave as in the
S=1/2 case, i.e., no energy gap and power-law decay of the spin correlation
function, while in the integer case the spectrum would have an energy gap
and the spin correlations would decay exponentially. Early numerical
work confirmed this conjecture\cite{exp1,num,moreo} and 
power-law decay of the spin correlation was found for half-odd-integer
spin, yet the exponent $\eta$ seemed to be S dependent.
When the powerful tools of conformal invariance started to be applied
to one-dimensional quantum systems it became clear that the integrable
S=1/2 Hamiltonian at low energies is equivalent to the SU(2) symmetric
Wess-Zumino-Witten (WZW) model with topological coupling 
constant $k=1$ \cite{affleck}. In this
model, $k$ is related to the central charge $c$ by $c=3k/(2+k)$ and the
critical exponent $\eta$ is given by $3/(2+k)$. It
was also found\cite{affleck} that a new class of integrable
antiferromagnetic Hamiltonians with arbitrary spin 
\cite{babu} were equivalent to the WZW model with $k=$2S. 

Based on
the SU(2) symmetry of the non-integrable h.o.i. spin Hamiltonians
Schulz\cite{schulz} suggested the following possible scenarios: \\ 
(a) The h.o.i. spin-S Heisenberg Hamiltonian belongs to the same universality
class as the spin-S integrable model. In this case both models would
have $c=3k/(2+k)$ and $\eta=3/(2+k)$ and $k=2S$. This scenario was
favored initially by Affleck \cite{affleck2}. \\
(b) The spin-S integrable models represent unstable multicritical
points in the phase diagram and the stable critical point is defined by
the S=1/2 Hamiltonian. Thus, the h.o.i. spin-S Heisenberg model is
expected to be equivalent to the $k=1$ WZW model having $c=1$ and
$\eta=1$, independent of the value of S. This position was favored by
Schulz and later on by Haldane and Affleck \cite{haldaf,affleck3}.

Since the derivations of scenarios a) and b) involve several assumptions
and approximations, unbiased numerical studies are needed to decide which one
is correct. The S=3/2 Heisenberg Hamiltonian is the simplest of the
higher half-odd-integer spin models. 
Since there are 4 degrees of freedom per
site the longest chains that can be exactly diagonalized with 
present computer capabilities have 14 sites. Previous numerical studies
conducted in such small chains gave conflicting
results \cite{moreo2,ziman}. Moreo\cite{moreo2}, analyzing the behavior of
spin correlation functions in lattices with $N\leq 12$ found
$\eta\approx 0.6$, while Ziman and Schulz\cite{ziman} obtained $\eta
\approx 1$ by studying the behavior of the lowest triplet and singlet
state energy gaps also in chains with $N\leq 12$. A Monte Carlo
calculation of the spin correlation 
function\cite{liang} in chains with up to 128 sites provided 
exponents in disagreement with the SU(2) WZW scaling theory.
The analysis of small systems is hampered by the presence of
logarithmic corrections which modify the power-law dependence of
various quantities in a significant way. 
In this
paper we will elucidate this problem numerically by studying 
sufficiently large
systems that cannot be solved with exact diagonalization, and
by Monte Carlo techniques only with considerable statistical error. 

For this purpose
we will use the density-matrix renormalization-group technique (DMRG) 
\cite{white}
keeping up to $m=1200$ states per block and considering chains up to 
$N=60$ sites.
For the largest systems considered we estimate that the ground state energy 
has an absolute error
of less than $7\times 10^{-6}$, by considering different $m$ values.
Our results for small systems, $N\le 14$,  agree with 
exact calculations \cite{parkinson,lin,moreo} and the ground state energy is
exact up to $N=12$.

To calculate the conformal central charge $c$ (or equivalently $k$)
we account for the finite-size corrections
of the ground-state energy per site, $E_0/N$, 
as derived from the WZW-theory\cite{affleck3}
\begin{equation}
\label{eq:c}
\frac{E_0}{N} =e_{\infty}-\frac{v \pi} {6 N^2}\left[\frac{3k}{2+k}+\frac{3k^2} 
{8\ln^3(BN)}
\right]-\frac{S}{N^4}.
\end{equation}
The coefficient $B$ depends on the effective
coupling constant (see below) and can be dropped only in the large-N 
limit\cite{horsch} . The last higher-order term reflects a 
finite size correction to the conformal charge.
In principle
Eq.(1) may be fitted with 5 free parameters, however we found that more
accurate results are obtained by an independent calculation of the 
spin velocity $v$ and of $B$.

The spin velocity can be determined
numerically from the excitation energy
$\Delta_q=E(2\pi/N)-E(0)=v\sin(2\pi/N)$, 
where $E(2\pi/N)$ is the energy of the
first excited state with the lowest non-zero momentum obtained
from the spin excitation spectrum \cite{karen} and the Davidson algorithm 
\cite{dav} (with estimated absolute error $\simeq 10^{-2}$).  
Studying the behavior of $\Delta_q/N$ vs $\sin(2\pi/N)/N$, as shown in
Fig.~1, to minimize the errors, 
we obtain a rather precise value for the spin velocity $v=3.87\pm 0.02$.
This implies an enhancement factor of 1.29 with respect to the spin-wave
result $v_{sw}=2S=3$. This factor 
is significantly lower than the value $\pi/2$ for the spin-1/2 chain
(where $v_{sw}=1$)
thus implying that the quantum renormalization of the low-lying  spin 
excitations is weaker.
The velocity is in good agreement with the measured spin-wave
velocity $v=1.26 v_{sw}$ for the spin-3/2 chain system
$CsVCl_3$\cite{itoh95}. 
The value reported for $AgCrP_2S_6$, $v=1.53
v_{sw}$\cite{mutka95} , appears too high in view of our results.   
Our $v$ is higher than 
the value $v=3.64\pm 0.08$ given by Yamamoto \cite{yamamoto} .
The latter was calculated
using a world-line Monte Carlo approach with larger error bars in the
excitation energies ($\sim 10\%$ as estimated from the figures).

\begin{figure}
\centerline{
\psfig{width=8.5truecm,bbllx=60pt,bblly=55pt,bburx=580pt,bbury=700pt,angle=-90,file={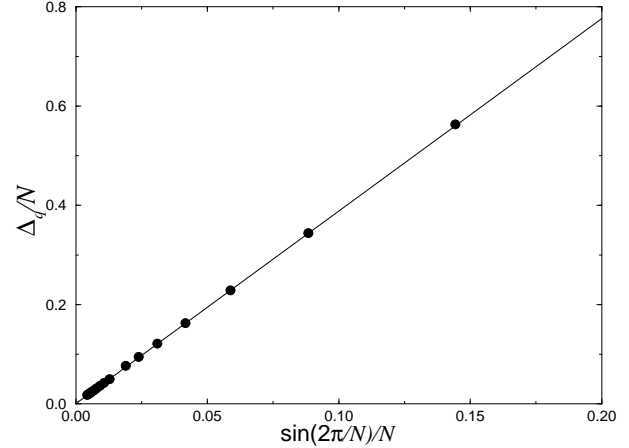}}}
\caption{Excitation energy $\Delta_q/N$ vs. $\sin(2 \pi/N)/N$. 
The linear interpolation yields the spin velocity (slope) $v=3.87\pm 0.02$
(full line). }
\label{fig:one}
\end{figure}

The value of the coupling constant $g$ can be determined  
directly by writing Eq.~1 in its equivalent form:
\begin{equation}
\label{eq:g}
\frac{E_0}{N} =e_{\infty}-\frac{v \pi} {6 N^2}\left[\frac{3k}{2+k}+
\frac{(2\pi g)^3}{\sqrt{3}k} 
\right] -\frac{S}{N^4},
\end{equation}
where
\begin{equation}
g(g_0,N,k)=\frac{g_0}{1+\frac{4\pi g_0}{\sqrt{3}k} \ln(\frac{N}{l_0})}
\end{equation}
and
\begin{equation}
\label{eq:b}
B=\frac{e^{g_{eff}^{-1}}}{l_0} \;\; ;\; g_{eff}=\frac{4\pi}{\sqrt{3}k} g_0.
\end{equation}
From Eqs. (2) and (3) 
and the velocity given above
we find $e_{\infty}=-2.82833(1)$, $k=1.00(1)$
and the coupling constant $g_0=g(l_0=20)=0.055(5)$. 
The error was estimated using different subsets of the data.
The constant $S$ depends on the smallest system included 
in the fit, {\it e.g.} for $N\ge 12$ we found $S=20$.
From Eq.~\ref{eq:b}, $B=0.6$ and $g_{eff}=0.4$.

Using the previous values of $v$ and $B$ and
from a three parameter fit of Eq.~\ref{eq:c} 
using $N\ge 8$
we obtain consistently $e_{\infty}=-2.82833$,
$k=1\pm 0.01 $ and
$S=15$ (Fig.~2).
Our ground-state energy is considerably lower than the energy
$e_{\infty}=-2.8248$ obtained in a recent Monte Carlo study
Ref.~\cite{sun}, where a $1/N^2$-scaling of the energy was assumed.
From our results for $k$ we conclude that $c=1.00\pm 0.01$

\begin{figure}
\leftline{
\psfig{width=8.5truecm,bbllx=60pt,bblly=55pt,bburx=580pt,bbury=700pt,angle=-90,file={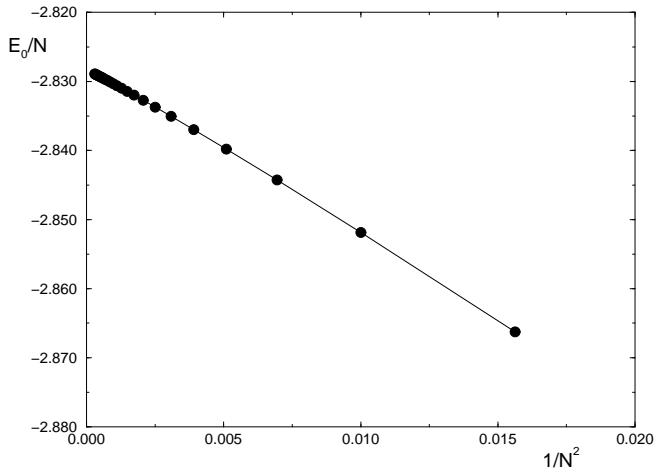}}}
\caption{
Scaling of the ground state energy $E_0/N$. 
Interpolation including logarithmic corrections yields the central
charge $c=1.00\pm 0.01$ and an extrapolated value $e_\infty=-2.82833$
(full line). }
\label{fig:one}
\end{figure}

To obtain the critical exponent we calculated the spin correlation function
keeping 1000 states per block. The absolute error for the largest system
is lower than $10^{-3}$. We found that the scaling behavior obeyed by
the spin $1/2$ chain \cite{horsch} is absent in this case and we
extrapolated the finite-size
data with a quadratic polynomial in $1/N^2$. The correlation 
function in the thermodynamic limit 
is shown in Fig.~3. We interpolated this correlation function
using the theoretical prediction \cite{affleck3}
\begin{equation}
\omega(l,\infty)=a \frac{\sqrt{\ln(Bl)}}{l^{\eta}},
\end{equation}
and obtained $a=2.06$, $B=0.56$ and $\eta=1.02$. 
This fit is presented
in Fig.~3 where we added the correlations for $S=1/2$ to show that in
both cases they have the same multiplicative logarithmic corrections,
the coupling constant being smaller for $S=1/2$ \cite{horsch}.
We note that 
$a=\omega(l_0,\infty) \sqrt{g_{eff}} l_0$ 
and $B$ follows from Eq.~\ref{eq:b}.
Using the value for
$g_{eff}$ given above and the extrapolated value for $\omega(l_0,\infty)$
we obtain $a=2.07$ and $B=0.6$.
Thus the parameters obtained from the correlation function are in
complete agreement, whithin error bars, 
with those obtained from the scaling of the
ground state energy.

\begin{figure}
\leftline{
\psfig{width=8.5truecm,bbllx=60pt,bblly=55pt,bburx=580pt,bbury=700pt,angle=-90,file={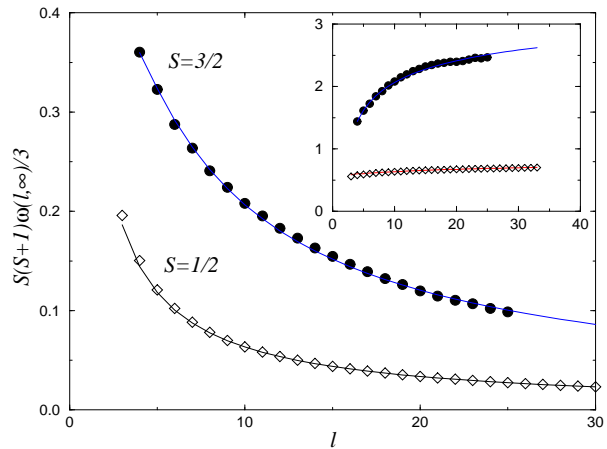}}}
\caption{Extrapolated and normalized spin correlation functions. The fit which
includes the multiplicative logarithmic correction gives $\eta=1.02$
(full line)
and coefficients that are in agreement with those obtained from the
scaling of the energy. The correlation function and fit of
the spin$-1/2$ chain\protect{\cite{horsch}} is shown for comparison. Inset:
Normalized correlation functions multiplied by $l$, showing the logarithmic
correction. }
\label{fig:one}
\end{figure}

On the other hand, if we assume $\eta=1$, we can analyze  
$l \omega(l,\infty)/l_0 \omega(l_0,\infty)=\sqrt{g(l_0)/g(l)}$.
This ratio directly reflects the dependence on
the coupling function \cite{horsch} and 
a single parameter fit yields $g(l_0=20)=0.052$.

The critical exponent $\eta$ can also be calculated from the scaling of 
singlet and triplet excitation energies \cite{affleck3} :
\begin{equation}
\label{eq:ei}
\Delta_i=\frac{E_i-E_0}{N}=\frac{2\pi v}{N^2}\left [\frac {\eta_i}{2}-
{{\bf S_L}\cdot {\bf S_R} \over \ln(BN)}\right ] + \frac{b_i}{N^2 \ln^2(BN)}.
\end{equation}
Here $\eta_i/2$  ($i=s,t$)  is the scaling dimension of the corresponding
 primary field,
and ${\bf S_L}$ and ${\bf S_R}$ are the spin operators related to the
 energy levels of the
conformal field theory describing the fixed point, with spin quantum numbers
 $s_L$ and $s_R$.
According to the WZW mapping
\begin{equation}
{\bf S_L} \cdot {\bf S_R}=\frac{1}{2}[s(s+1)-s_L(s_L+1)-s_R(s_R+1)],
\end{equation}
where $s$ is the total spin number. The triplet state $s=1$ and the 
first singlet excitation $s=0$ (for both cases $s_L=s_R=1/2$) become
degenerate in the thermodynamic limit. 
Therefore, if we take 
the appropriate linear combination of Eq.~\ref{eq:ei} for these excitations
the leading logarithmic corrections are cancelled out:
\begin{equation}
\label{eq:delta}
\Delta=\frac{1}{4N}\left[(E_s-E_0)+3(E_t-E_0)\right]=\frac {\pi v \eta}{N^2}
+ \frac{b}{N^2 \ln^2(BN)}.
\end{equation}
This quantity has been used before 
in Ref.~\cite{ziman} to estimate $\eta$ using three system sizes, 
{\it i.e.}
$N=8,10$ and $12$ sites with the result $\eta\simeq 1$. We reached  an 
excellent interpolation of our data (including a $S'/N^4$ term),
shown in Fig.~4, leading to $\eta=1.018$, $b=b_s-3b_t=-5.4$ and $S'=80$.

\begin{figure}
\leftline{
\psfig{width=8.5truecm,bbllx=60pt,bblly=55pt,bburx=580pt,bbury=700pt,angle=-90,file={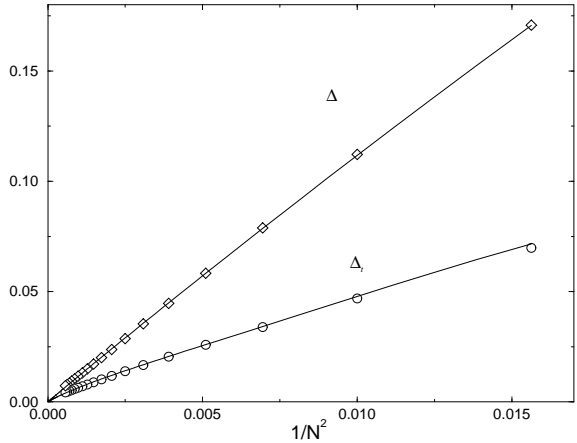}}}
\caption{Scaling of the triplet gap (circles) and the combined gap
$\Delta$ (diamonds). Equations (6) and (8) were used to extract the 
exponent $\eta$ (full lines).}
\end{figure}

We also obtained a good fit for the triplet finite-size gap ($\Delta_t$),
also shown in Fig.~4, but we were forced to add the subleading terms
($\sim (\ln\ln(BN))^2/\ln^4(BN)$ and $S''/N^4$) to obtain convergence.
In this case our results are $\eta=0.977$, $b_t=34$ and $S''=83$. For the
interpolation we have used the values for the velocity and $B$ calculated 
earlier.

In conclusion, we have found very accurate values for
the spin velocity, the central charge $c$ and the critical
exponent $\eta$ for the Heisenberg Hamiltonian with S=3/2 using the
DMRG technique that allowed us to study sufficiently large chains 
with negligible error. 
We found $c=1.00\pm 0.01$ and $\eta=1.02\pm0.02$ which shows, 
within error bars, that this model
belongs to the same universality class as the S=1/2 Heisenberg model and
the WZW model with $k=1$.
The $k=2S=3$ WZW model with $c=9/5$ and $\eta=3/5$ can be ruled out. 
These values have been obtained by an analysis of the ground-state energy,
the lowest singlet and triplet excitation energies 
as well as the spin correlation function. We have found
an excellent agreement between the parameters deduced from these
different data sets. 
Apart from a larger coupling constant the correlation function
of the spin-3/2 chain
has the same multiplicative logarithmic corrections as for
spin 1/2. 

We want to thank I.~Affleck, A. Muramatsu and I. Peschel
 for useful comments and H. Scherrer for computational support.
 This work was supported
in part by ONR under N00014-93-0495 and NSF under DMR-95-20776.

\end{document}